\documentclass[%
 aip,
 amsmath,amssymb,
 groupedaddress,
 reprint,%
]{revtex4-1}

\usepackage{graphicx}
\usepackage{dcolumn}
\usepackage{bm}

\usepackage{blindtext}
\usepackage{amsmath}
\usepackage{amsfonts}

\usepackage[dvipsnames]{xcolor}

\usepackage{physics}

\usepackage{blindtext}

\usepackage{lipsum}

\newcommand{\veck}{\mathbf{k}}
\newcommand{\vecq}{\mathbf{q}}

\newcommand{\tri}{\mathrm{t}}
\newcommand{\xx}{\mathrm{x}}
\newcommand{\ee}{\mathrm{e}}
\newcommand{\hh}{\mathrm{h}}
\newcommand{\FF}{\mathrm{F}}

\begin{document}



\title{Microscopic model of the doping dependence of line widths in monolayer transition metal dichalcogenides}


\author{Matthew R. Carbone}
\affiliation{Department of Chemistry, Columbia University, New York,
New York 10027}
\author{Matthew Z. Mayers}
\affiliation{Department of Chemistry, Columbia University, New York,
New York 10027}
\author{David R. Reichman}
\email{drr2103@columbia.edu}
\affiliation{Department of Chemistry, Columbia University, New York,
New York 10027}


\date{\today}

\begin{abstract}
A fully microscopic model of the doping-dependent exciton and trion line widths
in the absorption spectra of monolayer transition metal dichalcogenides in the
low temperature and low doping regime is explored. The approach is based on
perturbation theory and avoids the use of phenomenological parameters. In the low-doping regime,
we find that the trion line width is relatively insensitive to doping levels while the exciton line
width increases monotonically with doping. On the other hand, we argue that the
trion line width shows a somewhat stronger temperature dependence. The magnitudes
of the line widths are likely to be masked by
phonon scattering for $T \geq 20$ K in encapsulated samples in the low doping regime.
We discuss the breakdown of perturbation theory, which should occur at relatively low doping
levels and low temperatures. Our work also paves the way towards understanding a variety of related scattering processes, including impact ionization and Auger scattering in clean 2D samples.
\end{abstract}


\maketitle

\section{Introduction}
Monolayer transition metal dichalcogenides (TMDCs) are quasi-two-dimensional
(2D) materials known to exhibit extraordinary physical phenomena.\cite{mak2010atomically,Splendiani2010} These
materials may be viewed as semiconducting analogs of graphene,\cite{Novoselov2004,Novoselov2005,loh2010}
and present with non-trivial optical, electronic, and, under some circumstances,
topological and superconducting properties.\cite{chernHeintzReview,berkelbachRechmanReview2018} 
Due to their unique characteristics, monoloayer TMDCs have been proposed for myriad practical
applications\cite{Cai2018} such as opto-electronics,\cite{chhowalla2013,flory2015,memaran2015,wi2015}
field-effect transistors\cite{wang2012} and digital logic gates.\cite{ovchinnikov2014,lembke2015}
Of particular fundamental interest is the nature of electron-hole complexes
such as excitons\cite{qiu2013optical,wu2015exciton} and trions\cite{Berkelbach2013,mak2013tightly} in TMDCs.
Due to the reduced screening in 2D systems, such stable carrier
complexes may have anomalously large binding energies,
with that of the exciton reaching $\sim0.5$
eV,\cite{cheiwchanchamnangij2012quasiparticle,qiu2013optical,chernikov2014exciton,he2014tightly} and that of the trion reported to be in the
range of 20--35 meV,\cite{Berkelbach2013, Chernikov2014, Plechinger2015, Currie2015, Mitioglu2013}
implying that trions are bound even at room temperature. These observations
indicate that monolayer TMDCs are unique systems for investigating the properties of strongly interacting
quasiparticles. In addition, they may provide unprecedented
experimental clarity concerning the nature of interactions between these electron-hole
complexes and phonons,\cite{moody2015,Selig2016} as well as with charge carriers and other quasiparticles.

A standard means of probing the nature of the interactions of excitons and
trions with other excitations is via the broadening of line widths in clean
samples with respect to control parameters such as the temperature or
carrier density. Intrinsic homogeneous quasiparticle (QP)
line widths\cite{moody2015} are generally obfuscated by inhomogeneous broadening 
due to the high level of static defects in processing.
However, recent work has led to the observation of very narrow QP line widths via the preparation of ultra-clean samples by both dry transfer methods and chemical vapor deposition,\cite{Ajayi2017,cadiz2017,shree2019high} and by the usage of non-linear
spectroscopy to extract the homogeneous line width from inhomogeneously-broadened
spectra.\cite{moody2015}
The optical interrogation of the exciton and trion line widths in these less
defective samples offers a unique opportunity to understand the mechanisms of the
2D exciton and trion scattering processes in quasi-2D systems.

There are many factors that affect line broadening in monolayer
TMDCs, most notably interactions with phonons (as controlled by temperature)
and interactions with other charge carriers (as controlled by doping).
At very low temperatures and near the charge-neutrality point,\cite{Selig2016} it is expected that the intrinsic homogeneous line width due to lifetime broadening
may be observed if the sample is clean enough. As temperature increases,
phonons begin to play a significant role and will eventually dominate
the line broadening process. The interaction of excitons with phonons has
been studied in some detail in TMDCs,\cite{christiansen2017phonon,Selig2016} 
and a variety of coupling motifs have
been elucidated experimentally and theoretically.

Additionally, the 
concentration of electrons as controlled by gating can alter
lines widths and line shapes in a non-trivial fashion.\cite{shinokita2019continuous,miyauchi2018evidence,carvalho2015symmetry,qiu2013optical}Studies which have investigated the electron density dependence
of optical line shapes in monolayer TMDCs from the standpoint of the
Fermi-polaron picture provide a means of describing optical line broadening as a function of doping.\cite{sidler2017fermi,li2018optical,chang2019,efimkin2017,klawunn2011fermi,schmidt2012fermi,parish2013highly}
Such many-body multiple scattering theories are
essential for properly describing the full range of doping-dependent
behavior, as the Fermi Golden Rule breaks down at
sizable doping levels. However, the use of graphene gating and clean
samples renders the investigation of the doping regime
close to the charge-neutrality point possible.\cite{lherbier2008charge}
Here, detailed microscopic Golden Rule-based calculations may be performed
which can provide new insights into the line broadening mechanisms. Motivated by
the aforementioned recent experimental works,
we follow this latter path to assess how the elastic scattering of excitons and
trions with free charge carriers may alter line widths of both ground and
excited state excitonic complexes in the low doping regime. In 
particular, we investigate the circumstances for which doping-related broadening
may compete with phonon-induced broadening, and we discuss the breakdown of
the perturbative approach as a function of temperature and
carrier density. The importance of our work extends beyond the description of
linewidths and is of relevance for describing scattering processes such as 
Auger recombination and impact ionization in TMDCs.

Our paper is organized as follows: We first present an outline of the
microscopic theory in Section~\ref{sec: methodology}, focusing on the electron-exciton
scattering calculation, which is discussed in Subsection~\ref{subsec: m:exciton}.
Calculations for the electron-trion scattering are similar to that of the exciton and
discussed (briefly) in Subsection~\ref{subsec: m:trion}. The low-temperature
results for the trion and exciton line widths, in addition to the
details of the model and limitations of the Golden Rule approach, are presented and
discussed in Section~\ref{sec: resdisc}. Finally, in Section~\ref{sec:conclusion},
we summarize our conclusions and discuss outlook and potential future work. Details not
contained in the main text are located in several appendices.

\section{Methodology \label{sec: methodology}}
In this section, the elastic (energy-conserving) scattering of electrons from both
excitons and trions described within the Fermi Golden Rule approximation.
Additionally, because we work at the Golden Rule level of theory, bound states in scattering
are not considered.
While such a treatment can only be valid at extremely low doping densities, recent synthetic work
using encapsulated samples points to a route to experimentally controlled access to
this regime. Furthermore, the use of the Golden Rule allows for a very detailed microscopic
description,\cite{Cohen2003} the limitations which will be discussed in the following sections.

\subsection{Electron-exciton elastic scattering \label{subsec: m:exciton}}
In order to facilitate the computation, we use a simple variational guess for
the exciton wave function
\begin{equation}\label{exciton wave function (real space)}
    \phi(r) = \sqrt{\frac{\pi}{2\lambda^2}}
    e^{-r/\lambda},
\end{equation}
where $r$ is the relative coordinate of the two-body
system. The optimal effective Bohr radius $\lambda$ is chosen to best match
the functional form of (\ref{exciton wave function (real space)}) to the
ground state of a Wannier exciton in a Rytova-Keldysh potential
\cite{rytova2018,keldysh1979coulomb} found using exact diagonalization.

The second-quantized form of the exciton-free electron scattering state is
\begin{equation}\label{exciton-free electron ket}
    \ket{\veck_\xx, \veck_\ee} = \sum_{\veck'}
    \phi_{\alpha_\xx \veck_\xx + \veck'}^*
    \psi_{\veck_\ee}^*c_{-\veck'}^\dagger d_{\veck_\xx + \veck'}^\dagger
    c_{\veck_\ee}^\dagger \ket{0},
\end{equation}
which is a direct product state of the free exciton and electron states,
$\ket{\veck_x} \otimes \ket{\veck_e}.$
The wave function
\begin{equation}\label{exciton wave function (k space)}
    \phi_{\veck} = \sqrt{\frac{8\pi\lambda^2}{A}}g(\lambda k)
\end{equation}
satisfies normalization $\sum_{\veck} \phi_\veck^2 \rightarrow
\frac{A}{(2\pi)^2}\int \dd^2k\phi_\veck^2 = 1$ and
is derived by performing an in-plane Fourier transform of
(\ref{exciton wave function (real space)}), where
$g(x) = [1 + x^2]^{-3/2},$
$c_\veck$ $(d_\veck)$ are electron (hole) annihilation operators for momentum index $\veck$,
$A$ is the in-plane area of the 2D material, and $\alpha_\xx = m_\ee/M_\xx$ is the ratio of the electron
and exciton effective masses (which manifests during the coordinate transform
to relative/center of mass coordinates). The free-electron wave function
$\psi_\veck \propto e^{-i\veck \cdot \mathbf{R}}$ characterizes an electron which may
exhibit free in-plane motion,
and together with the center of mass coordinate of the exciton,
contributes only a global phase factor which may be ignored in subsequent
calculations, as it does not contribute to the determination of the scattering rate.

Scattering matrix elements are computed by evaluating the coupling between an
initial QP-free electron state, $\ket{\veck_\xx, \veck_\ee},$ and a final
QP-free electron state in which momentum $\vecq$ is transferred, $\bra{\veck_\xx +
\vecq, \veck_\ee - \vecq}.$ The second-quantized, momentum-conserving potential
energy operator $V = V_{\ee \hh} + V_{\ee \ee}$ that mediates this coupling may be
split into electron-hole and electron-electron components,
\begin{subequations}\label{potential energy operator eh}
\begin{equation}
    V_{\ee\hh} = -\sum_{\substack{\veck_1, \veck_2, \vecq \\ s = \uparrow,\downarrow}}v_\vecq c_{\veck_1+\vecq}^{s\dagger}
    d_{\veck_2-\vecq}^\dagger d_{\veck_2} c_{\veck_1}^{s},
\end{equation}
and
\begin{equation}\label{potential energy operator ee}
    V_{\ee\ee} = \frac{1}{2}\sum_{\substack{\veck_1, \veck_2, \vecq \\ s_1, s_2= \uparrow,\downarrow}} v_\vecq c_{\veck_1+\vecq}^{s_1\dagger}
    c_{\veck_2-\vecq}^{s_2\dagger} c_{\veck_2}^{s_2} c_{\veck_1}^{s_1},
\end{equation}
\end{subequations}
where $v_\vecq = \frac{2\pi e^2}{A q \varepsilon(q)}$ is
the magnitude of the two-body interactions and $\varepsilon(q)$ is a static dielectric function
discussed in Section~\ref{subsection: dielectric function}. The exciton-free electron elastic scattering matrix elements
are henceforth defined as 
\begin{equation}\label{exciton-free electron SME def}
V(\vecq, \veck_\ee, \veck_\xx)
= \mel{ \veck_\xx + \vecq, \veck_\ee - \vecq}{V}{\veck_\xx, \veck_\ee}.
\end{equation}

Once matrix elements have been computed, the line width $\Gamma(n; \veck_\xx)$ is
calculated by summing over all final exciton states,
\begin{equation}\label{fermi's Golden Rule}
    \Gamma(n, \veck_\xx) = \frac{\hbar A}{(2\pi)^2} \int \dd^2 q \: w(\vecq; n, \veck_\xx).
\end{equation}
Here, $w(\vecq; n, \veck_\xx)$ is a partial scattering rate computed for fixed momentum transfer using Fermi's
Golden Rule,
\begin{multline}\label{scattering rate}
    w(\vecq; n, \veck_\xx) = \frac{2\pi}{\hbar}
    \sum_{\veck_\ee} |V(\vecq, \veck_\ee, \veck_\xx)|^2 f(k_\ee)[1 - f(|\veck_\ee - \vecq|)] \\
    \times \delta \left(\frac{\hbar^2 k_\xx^2}{2M_\xx} + \frac{\hbar^2k_\ee^2}{2m_\ee}
    - \frac{\hbar^2|\veck_\xx + \vecq|^2}{2M_\xx} - \frac{\hbar^2|\veck_\ee-\vecq|^2}{2m_\ee} \right),
\end{multline}
where the Fermi-Dirac distribution
\begin{equation}
    f(k) = \left[e^{(\hbar^2k^2/2m_\ee - \mu)/k_\mathrm{B}T} + 1\right]^{-1}
\end{equation}
contains doping-density ($n$) dependence through the chemical potential 
$\mu = k_\mathrm{B}T\ln \left[ \exp{\varepsilon_\FF/k_\mathrm{B}T} - 1\right],$
and the Fermi energy of a 2D electron gas, $\varepsilon_\FF = \pi \hbar^2 n/m_\ee.$
In order to simplify the calculations, the parameter $\veck_\xx = 0$ is taken in all
computations, effectively choosing a reference frame in which the exciton is at rest. 
For further details, we refer the reader to Ref.~\onlinecite{Cohen2003}, where similar
calculations are performed for anisotropic 3D systems.

\subsection{Electron-trion scattering \label{subsec: m:trion}}
Computation of the trion-free electron elastic scattering line width
contribution is similar to that of the excitonic case in all ways except for the
determination of the scattering matrix elements. The trion-free electron
scattering state is constructed similarly to that of (\ref{exciton-free
electron ket}), with a few key distinctions to be noted below. Explicitly, we write
this scattering state as,
\begin{multline} \label{trion-free electron ket}
    \ket{\veck_\tri, \veck_\ee} = \sum_{\substack{\veck_1, \veck_2 \\
    s_1, s_2, s_\ee}}
    \xi_S^*(s_1, s_2)\Phi_{\alpha_\tri \veck_\tri + \veck_1,\alpha_\tri\veck_\tri + \veck_2}^*\psi_{\veck_\ee}^*\\ \times
    c_{-\veck_1}^{s_1\dagger}c_{-\veck_2}^{s_2\dagger}
    d_{\veck_\tri + \veck_1 + \veck_2}^\dagger
    c_{\veck_\ee}^{s_\ee\dagger} \ket{0}.
\end{multline}
Note the introduction of a spin wave function which constrains the
trion to the singlet spin configuration,
$\xi_S(s_1, s_2) =  \bra{s_1 s_2}\ket{S},$
via the projection of a two-fermion spin state $\bra{s_1s_2}$
on the singlet state $\ket{S}.$
The projection satisfies the properties,
$\sum_{s_1, s_2} \xi_S^*(s_1, s_2) \xi_S(s_1, s_2) = \bra{S}\ket{S} = 1,$
and $\xi_S(s_1, s_2) = - \xi_S(s_2, s_1)$
as per Fermionic anti-commutation rules. Given that the trion triplet state is, at most,
weakly bound, we only consider only singlet to singlet scattering.

The trion wave function, $\Phi,$ is given by
\begin{equation}\label{trion wave function}
    \Phi_{\veck_1, \veck_2} = \mathcal{N}\frac{8\pi\lambda_1\lambda_2}{A}
    g(\lambda_1k_1)g(\lambda_2k_2).
\end{equation}
Here, $\lambda_1$ and $\lambda_2$ are variational parameters associated with
the Chandrasekhar-type wave function,\cite{Berkelbach2013,chandrasekhar1944} and
the constant $\mathcal{N}$ is a normalization factor 
\begin{equation} \label{trion energy normalization factor and kappa}
    \mathcal{N} = \frac{1}{\sqrt{1 + \kappa^2}}, \quad
    \kappa = \frac{4 \lambda_1 \lambda_2}{(\lambda_1 + \lambda_2)^2}
\end{equation}
which arises during the variational minimization of the trion
binding energy.\cite{Sandler1992}

Once the matrix elements 
\begin{equation} \label{trion mel}
    \mathcal{V}(\vecq, \veck_\ee, \veck_\tri) =
    \mel{\veck_\tri + \vecq, \veck_\ee - \vecq}{V}{\veck_\tri, \veck_\ee}
\end{equation}
are computed, the trion line width may be determined using
(\ref{fermi's Golden Rule}) and (\ref{scattering rate}) in the same way
as for the exciton case (with the appropriate substitutions, e.g. the initial
QP momentum $\veck_\xx \rightarrow \veck_\tri$, the mass ratio $\alpha_\xx \rightarrow
\alpha_\tri = m_\ee/M_\tri$, etc.). Line widths for low doping
densities are reported in Section~\ref{sec: resdisc}, computational details
of this calculation are given in Appendix~\ref{apdx: computational} and
the physical parameters used may be found in the caption of Fig~\ref{fig:res}.

\subsection{Dielectric Function \label{subsection: dielectric function}}
The dielectric function $\varepsilon(q, \omega)$ takes into
account properties of the monolayer TMDC, the surrounding medium,
and the excess electron gas,\cite{Stern1967} respectively,
and may be broken down into distinct contributions as\cite{tim_unpublished}
\begin{equation}
    \label{improved dielectric function}
    \varepsilon(q, \omega) = \varepsilon_\mathrm{I}(q) + \varepsilon_\mathrm{II}(q, \omega).
\end{equation}
We follow previous work~\cite{giuliani1982lifetime} and screen the direct and
exchange interactions, in contrast to the usual Bethe-Salpeter treatment of bound state
formation where the exchange interaction is unscreened.\cite{strinati1988application,rohlfing2000electron,hanke1979many,benedict2002screening}
The first term consists of a static contribution from the monolayer TMDC and
surrounding layers in the absence of doping,
\begin{equation}\label{epsilon1}
    \varepsilon_\mathrm{I}(q) = \varepsilon_0(1 + 2\pi \chi_\mathrm{2D}q),
\end{equation}
where $\varepsilon_0 = (\varepsilon_a + \varepsilon_b)/2$ is the dielectric constant of the surrounding
medium\cite{rytova2018,keldysh1979coulomb}
(the average of the two encapsulating dielectrics) and
$\chi_\mathrm{2D}$ is the dielectric polarizability of the 2D
material.

The second term is due to the presence of doping electrons and is generally frequency
dependent. We follow Stern\cite{Stern1967} and treat the excess electrons as a 2-dimensional
homogeneous electron gas (HEG). In the static ($\omega = 0$) approximation, this yields
\begin{equation}\label{epsilon2-static}
    \varepsilon_\mathrm{II}(q, 0) = \frac{2 m e^2}{\hbar^2q}
    \begin{cases}
    1 & \text{if } q \leq 2k_\FF \\
    1 - \sqrt{1 - (2k_\FF/q)^2} & \text{if } q > 2k_\FF
    \end{cases}.
\end{equation}
Note that (\ref{epsilon2-static})
implicitly carries a doping density ($n$) dependence through
the Fermi momentum $p_\FF = \hbar k_\FF = \hbar \sqrt{2\pi n}.$

To motivate this choice, we observe that $\varepsilon(q, 0)$ captures the correct
behavior in both the small-wavelength and low-doping limits.
In the low-doping limit, the Stern-like term vanishes and the dielectric function
$\varepsilon(q, 0) \rightarrow \varepsilon_\mathrm{I}(q),$
which is the dielectric function of the material and its surroundings. The
low $q$-limit suppresses the term containing the polarizability and diverges like $1/q,$
correctly screening the 2D Coulomb interaction at small $q$.\cite{glazov2018breakdown}

If doping levels are large enough, the static approximation presented
above will fail.\cite{glazov2018breakdown,scharf2019dynamical} Although this signals one aspect of the high-doping density breakdown of the Golden Rule, one way to potentially extend its domain of
validity of is to utilize a frequency-dependent scattering matrix element as discussed in Ref.~\onlinecite{giuliani1982lifetime}. This leads to a
dielectric function derived from the 2D Lindhard function,
$\varepsilon_\mathrm{II}(q, E_\mathrm{eff}/\hbar),$\cite{giuliani1982lifetime,Stern1967} evaluated at the effective energy 
\begin{equation} \label{eff energy}
    E_\mathrm{eff} \equiv E(k_\ee) - E(\abs{\veck_\ee - \vecq}) 
    = \frac{\hbar^2(2\mathbf{k}_\ee \cdot \mathbf{q} - q^2)}{2m_0},
\end{equation}
which is the energy difference between the initial and final states of the
scattering electron. The details of $\varepsilon_\mathrm{II}(q, \omega)$ are presented in
Ref.~\onlinecite{Stern1967} and in Appendix~\ref{rpa pol}.

\section{Results and Discussion \label{sec: resdisc}}
The Fermi Golden Rule is expected to be valid only in the ultra-low doping
regime ($\varepsilon_\FF \ll \varepsilon_\tri \sim 10^{12} \mathrm{cm}^{-2}$), where $\varepsilon_\tri$ 
denotes the trion binding energy in the limit of zero doping. As the doping level increases,
many-body, multi-scattering effects become prominent,\cite{Chang2018-arXiv}
and a Fermi-polaron-like picture appears to be
required.\cite{efimkin2017,demler2017,chang2019} 
Since the low doping regime is now potentially controllable and
accessible in encapsulated samples with graphene gating layers, a
Golden Rule approach is useful in enabling a fully microscopic treatment in this
restricted regime.

\begin{figure}[!htb]
\includegraphics[width=\columnwidth]{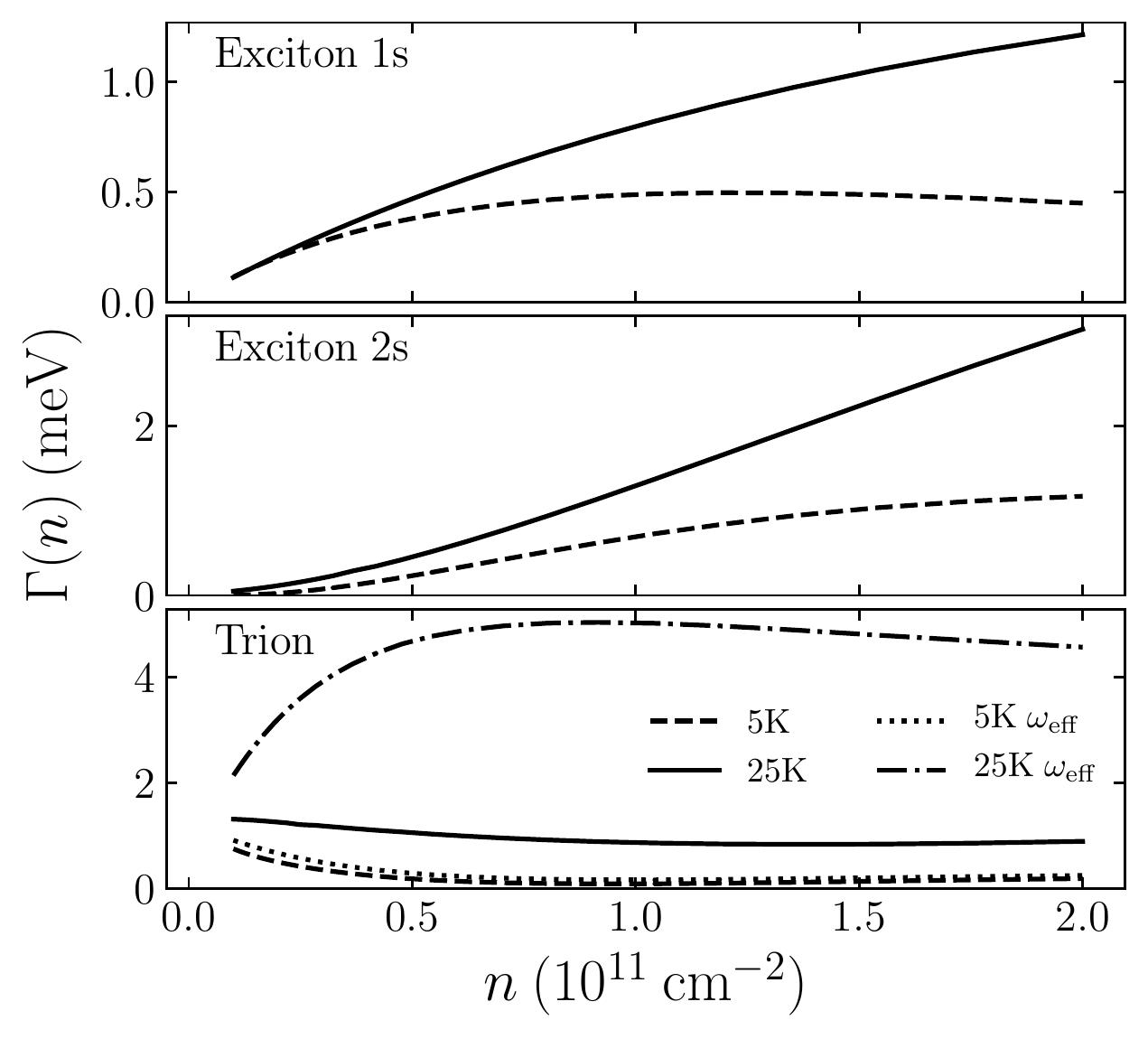}
\caption{\label{fig:res} Line width broadening of monolayer MoSe$_2$ as a function of
electron doping density for BN-encapsulated ($\varepsilon_0 = 4.5$)\cite{Gielisse1967}
monolayers. The following parameters
were used: in the exciton calculation, the effective Bohr radii $\lambda_0 = 10.3$ and in
the case of the trion, $\lambda_1=\lambda_0$ and $\lambda_2 =
25.2$~\AA.\cite{Berkelbach2013} In the exciton 2s elastic scattering, $a=7.79$~\AA~and 
$b=6.20$~\AA~(see Appendix~\ref{2s apdx}). The electron (hole) effective masses employed were $0.49$
$(0.61)$ (in units of $m_0$),\cite{Rasmussen2015} and the polarizability $\chi_\mathrm{2D} = 8.23$~\AA.\cite{Berkelbach2013} 
In the case of exciton elastic scattering, the singlet and triplet contributions are identical as
the exchange contribution to the potential dominates; trion triplet states are not considered.
Additionally, screening using the effective frequency-dependent dielectric function (see Eq.~\ref{eff energy}) are presented for the trion, as the effective screening does not appear to affect the exciton line width.}
\end{figure}

Line widths versus doping level for both the exciton and trion lines
are displayed for 5 and 25~K in Fig.~\ref{fig:res}. Results are
presented for the experimentally-relevant
case of a layer encapsulated by dielectric media with properties
mimicking that of boron nitride. 
We also note that in a hypothetical suspended sample ($\varepsilon_0 = 1$), $\Gamma$ is
enhanced compared to results presented in Fig.~\ref{fig:res} (e.g. roughly 5 meV at $10^{11}$ cm$^{-2},$ compared to only 1 meV in the encapsulated case) and is comparable, or even larger than that associated with phonon-induced broadening, since the scaling of
$\Gamma$ with respect to the background
dielectric function varies roughly as $\varepsilon_0^{-2}.$
It should also be noted that in experiments the encapsulating
layers are of finite thickness, and while this situation can be handled
theoretically,\cite{andersen2015,cavalcante2018} we do not do so here
as it complicates the treatment of the dielectric function.
We thus expect the true magnitude of line width values to be somewhat larger than the values presented in Fig.~\ref{fig:res}. Additionally, while we have
also carried out an investigation of inelastic electron-capture scattering,
we find that elastic scattering dominates the line widths in the regimes we
consider. Thus, we only focus on the elastic scattering contributions.

We first discuss trion line broadening. For doping levels $n > 0.4 \times
10^{11}$ cm$^{-2},$ the trion line width in all cases is largely independent of doping
density. The upturn seen in the static screening trion line width as
doping density decreases is likely an artifact of behavior embedded in the function
$\varepsilon_2(q).$ Indeed, a  suppression of the $q^{-3}$ behavior for large
$q$ of this function leads to an essentially flat trion line width as a function
of doping level, similar to that seen in Fermi-polaron-like theories and in some
experiments.\cite{astakhov2000oscillator,smolenski2018shubnikov} It should
be noted that in these approaches, however, the trion line broadening is
controlled by a phenomenological input parameter.\cite{efimkin2017} 
Here, our fully microscopic approach allows for the microscopic extraction of the magnitude and
temperature dependence of the trion line width. While the static and
effective frequency-dependent screening cases are largely in agreement at low $T,$
the same cannot be said for results at 25K. Given the subtle changes in the
scattering matrix elements except at small $q,$ this difference likely arises from
the larger accessible density of states available at higher densities away from
$\omega = 0$ in the screening function.

\begin{figure}[!htb]
\includegraphics[scale=0.32]{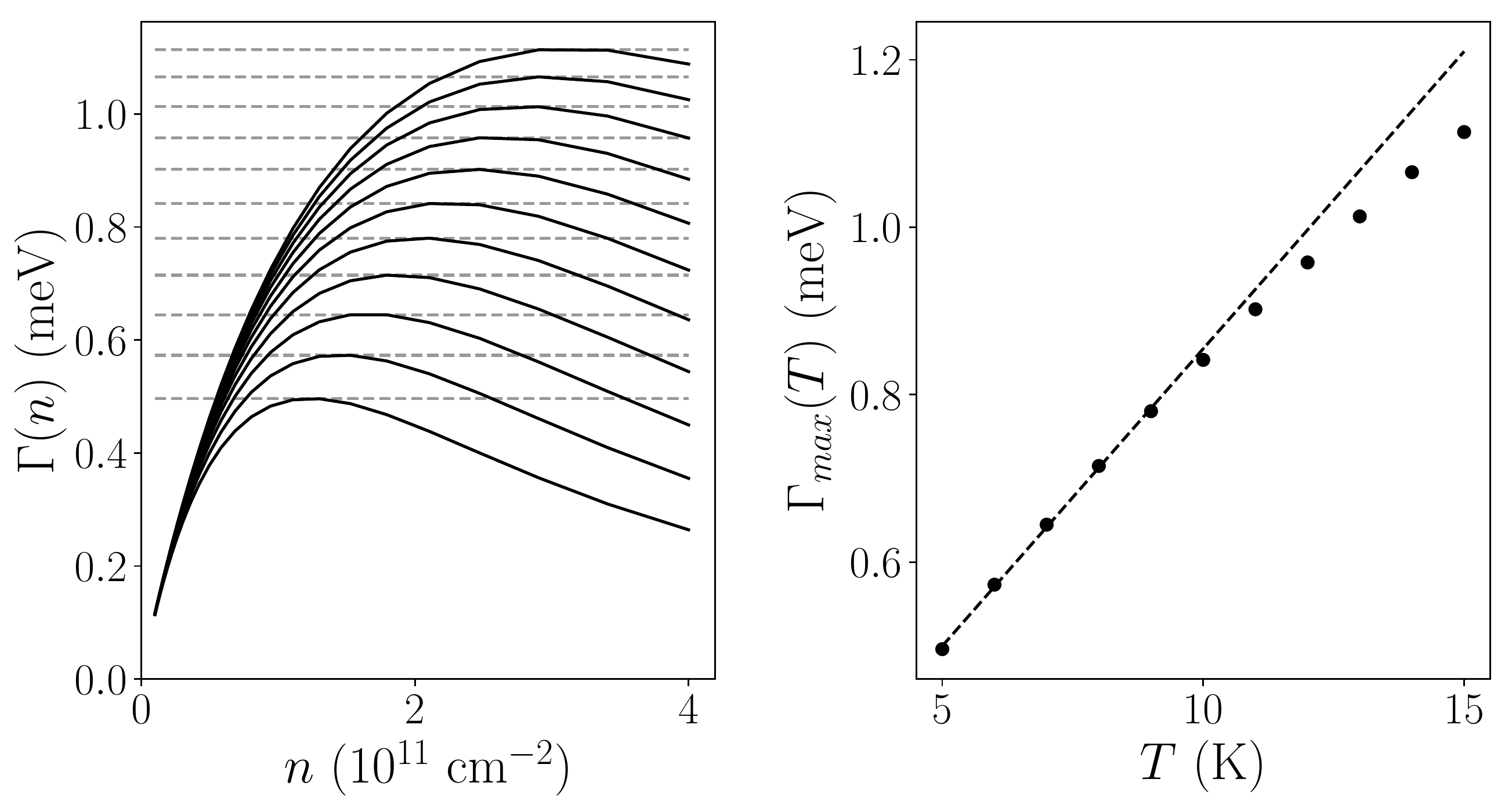}
\caption{\label{fig:resT} Doping dependence of the 1s exciton line
width at temperatures $T= 5, 6, ..., 15$ K (left). Parameters describe
monolayer MoSe$_2$, as seen in Fig~\ref{fig:res}. Lower line widths
correspond to lower temperatures. The horizontal
dashed lines show the plateau location. The value of the maximum (plateau) is also
plotted as a function of temperature, showing clear linear behavior at low $T$ (right).} 
\end{figure}

We now turn to the broadening of the exciton line. Unlike the trion case,
the exciton line width monotonically increases as a function of doping
density at low values of $n$ in the 25 K case. This is again in agreement
with experimental expectations as well as the behavior found in many-body
approaches.\cite{efimkin2017,demler2017,Chang2018-arXiv} In particular,
in these latter theoretical approaches, an approximately linear dependence of the line
width on doping manifests over a much wider doping density range for the exciton
line. The very same behavior arises from the
Golden Rule at extremely low doping. The decrease of the
slope of the line width as $n$ increases, most clearly demonstrated in the
near-plateau of the 5 K exciton line widths above $n=0.8 \times 10^{11}$~cm$^{-2},$ is a signature of the breakdown of the Golden Rule. Specifically, due
to the $\varepsilon_\FF/k_\mathrm{B}T$ term in the Fermi-Dirac distribution function, the crossover from the non-degenerate to the degenerate electron gas limit will induce a change in the doping dependence of the excitonic line width from a linear scaling $\Gamma \sim n$ at low doping to an eventual
plateau $\sim k_\mathrm{B}T,$ and then an unphysical decline with increasing $n.$
This same trend is reported in Ref.~\onlinecite{Cohen2003} for the quantum well case.
We systematically examine this behavior in
Fig.~\ref{fig:resT}, which shows the doping and temperature dependence of this
behavior. If one focuses on the more physically-described regime $n < 0.8 \times 10^{11}$~cm$^{-2},$ it is observed that, unlike in the trion case, doping-induced exciton line broadening is
largely insensitive to temperature variations in the range $T=5$-$25$~K. Furthermore,
given the fact that phonon-induced line broadening is suppressed at these
temperatures, doping induced line broadening effects may be observable at $T=5$ K
in clean, encapsulated samples even for doping densities as low as $n \sim 2
\times 10^{11}$~cm$^{-2},$ especially with respect to the 2s line, where the line broadening effects appear to be slightly enhanced compared to the ground state. 

\section{Conclusion}\label{sec:conclusion}
In this work we have employed perturbation theory to calculate the rates of
electron-exciton and electron-trion scattering in monolayer TMDCs in the low doping
density limit. Our approach is fully microscopic with respect
to all input parameters and functions, including matrix elements and the dielectric
screening model. On the other hand, it is expected that the Fermi Golden Rule should
break down at low doping densities close to the degeneracy crossover of the
electron gas in the monolayer, and some caution must be exercised with respect to the
use the forms of the dielectric screening functions employed here.\cite{glazov2018breakdown}
Avoiding these approximations allows for the description of a much broader range of
doping, but requires a full frequency-dependent many-body
treatment.\cite{efimkin2017,demler2017,chang2019}

Accepting the above limitations, the calculations presented here still allow for some
important conclusions to be drawn. First, we find that with a reasonable treatment
dielectric environment, exciton line widths arising from exciton-electron scattering 
on the order of 1~meV or higher are
possible at low temperatures in the low doping regime accessible in
encapsulated, graphene-gated samples. Thus, even mild doping may provide a line
broadening mechanism that can compete with (but not necessarily exceed) lifetime and
phonon-related broadening in this regime. As expected from previous many-body
calculations in the very low doping regime, 
the growth of the excitonic line width is monotonic with increasing
$n$, while the trion line width is largely insensitive to doping. However we find
that the trion line width is sensitive to temperature variations even over the small
range $T=5$-$25$~K, a somewhat unexpected feature from the standpoint of many-body
theories such as that of Ref.\cite{efimkin2017} where the trion line width is partly described
by a phenomenological input parameter. Lastly, we find that excited state exction line
broadening is somewhat larger and shows more sensitivity to increases in doping levels. Future
work should be devoted to testing the veracity of these predictions
and to understand how they merge with many-body approaches which have been
applied to study the higher doping density regime.\cite{fey2019theory}

In conclusion, we have provided a microscopic model for understanding how the
scattering of excitons and trions surrounded by an electron gas in monolayer TMDCs
may induce line broadening in the very low doping density limit at low temperatures.
A more detailed effort aimed at placing these contributions in the context of other
mechanisms, such as exciton-phonon scattering, is worth of future study. 
In addition, the approach adopted here
may be of use for the calculation of the rates of processes such as Auger
recombination\cite{huang2006quantized,wang2006auger,hu2016slow} in
dimensionally-confined systems. These and related topics will the subject of future
investigations.

\begin{acknowledgments}
M.R.C. acknowledges support from the United States Department of Energy through the Computational Sciences
Graduate Fellowship (DOE CSGF) under grant number: DE-FG02-97ER25308. D.R.R. acknowledges support by 
NSF-CHE 1464802. The authors acknowledge fruitful discussions with Ian S. Dunn, Timothy C. Berkelbach, Guy
Ramon, Archana Raja, Alexey Chernikov and Mikhail Glazov.

\end{acknowledgments}

\appendix

\section{RPA Polarizability} \label{rpa pol}
Following the definition in Stern,\cite{Stern1967} in this appendix we present the frequency-dependent 2D electron gas
polarizability, $\chi,$ and its $\omega \rightarrow 0$ limit. The general
form of $\chi$ is $\chi(q, \omega) = \chi_1(q,\omega) + i \chi_2(q, \omega),$ where
\begin{widetext}
\begin{equation}
    \chi_1(z, \tilde{u}) = \frac{e^2m}{\hbar^2q^2\pi}\left\{1 - \frac{C_-(z, \tilde{u})}{2}
    \sqrt{\left(1 - \tilde{u} \right)^2 - z^{-2}} - \frac{C_+(z, \tilde{u})}{2}
    \sqrt{\left(1 + \tilde{u} \right)^2 - z^{-2}} \right\}
\end{equation}
and
\begin{equation}
    \chi_2(z, \tilde{u}) = \frac{e^2m}{\hbar^2q^2\pi}\left\{ \frac{D_-(z, \tilde{u})}{2} \sqrt{ z^{-2} - \left(1 - 
    \tilde{u} \right)^2}
    - \frac{D_+(z, \tilde{u})}{2} \sqrt{ z^{-2} - \left(1
    + \tilde{u} \right)^2}
    \right\},
\end{equation}
\end{widetext}
where $z \equiv q/2k_\FF$ and $\tilde{u} \equiv  2\omega m /\hbar q^2.$
Note that the quantities in the braces, $\{\cdot\},$ are dimensionless.
The functions $C$ and $D$ are defined as follows,
\begin{equation}
    C_\pm(z, \tilde{u}) \equiv \begin{cases} 
    (z\pm \tilde{u})/|z \pm \tilde{u}| & \text{if } |z \pm \tilde{u}| > 1 \\
    0 & \text{otherwise}
    \end{cases}
\end{equation}
and
\begin{equation}
    D_\pm(z, \tilde{u}) \equiv \begin{cases} 
    0 & \text{if } |z \pm \tilde{u}| > 1 \\
    1 & \text{otherwise}
    \end{cases}.
\end{equation}
In the static approximation we note that $\chi_2(q, 0) = 0$
and $\chi_1$ reduces to (\ref{epsilon2-static}), where generally
\begin{equation}
    \varepsilon_2(q, \omega) = 2 \pi B(q, \omega) \chi(q/2k_\FF, 2\omega m/\hbar q^2),
\end{equation}
and $B(q, \omega) = \sqrt{q^2 - \varepsilon_0 \omega^2 c^{-2}}.$

\section{Exciton-electron elastic scattering}
In this appendix, we outline the details of the $X + e^- \rightarrow
X + e^-$ scattering calculation, including accounting for electron spin. Here, and in
Appendix~\ref{apdx trion}, we follow closely with the approach of Ref.~\onlinecite{Cohen2003},
generalizing to the strict 2D limit and filling in necessary details.

In the following, it will be useful to keep in mind the electron and
hole anti-commutation relations
$$
\{d_{\veck}^{s\dagger}, c_{\veck'}^{s'\dagger}\} = \{d_{\veck}^{s\dagger},
c_{\veck'}^{s'}\} =
\{d_{\veck}^{s}, c_{\veck'}^{s'\dagger}\} = \{d_{\veck}^{s},
c_{\veck'}^{s'}\} = 0,$$
(electrons and holes always anti-commute) and,
$$
\{x_{\veck}^s, x_{\veck'}^{s'}\} = \{x_{\veck}^{s\dagger},
x_{\veck'}^{s'\dagger}\}
= 0; \quad
\{x_{\veck}^s, x_{\veck'}^{s'\dagger}\} =
\delta_{\veck\veck'}\delta_{ss'},$$
where $x = c, d.$ Also, recall that $\psi_{\veck_\ee}$ ends up as a 
global phase factor in the expression for the scattering rate, and will
be ignored in the following derivations.

\subsection{General form of the matrix elements}
A prudent first step to
computing (\ref{exciton-free electron SME def}) is to split up $V$ into
its constituent parts and evaluate them independently on the initial state
$\ket{\veck_\xx^\alpha, \veck_\ee^\beta},$ where spin indexes have been added
as superscripts (the exciton spin references the electron; hole spin will
not be important). In the case of the electron-hole component, we have
\begin{equation}\label{V_eh on exciton, with spin}
\begin{split}
    V_{\ee \hh}\ket{\veck_\xx^\alpha, \veck_\ee^\beta} &= 
    \sum_{\veck'\vecq'}v_{q'}\phi^*
    \underbrace{
    c_{-\veck' + \vecq'}^{\alpha \dagger} d_{\veck_\xx + \veck' - \vecq'}^\dagger}_\text{self-interaction}
    c_{\veck_\ee}^{\beta \dagger}\ket{0} \\
    &-  \sum_{\veck'\vecq'}v_{q'} \phi^*
    c_{-\veck'}^{\alpha \dagger} d_{\veck_\xx + \veck' - \vecq'}^\dagger
    c_{\veck_\ee + \vecq'}^{\beta \dagger}\ket{0},
\end{split}
\end{equation}
where $\phi^* \equiv \phi^*_{\alpha_\xx\veck_\xx + \veck'}.$
The first term in the above equation only contains information about the exciton
interacting with itself (self-interaction) and is therefore discarded.
The electron-electron component is calculated in a similar fashion and does
not contain self-interaction terms:
\begin{equation}
    V_{\ee\ee}\ket{\veck_\xx^\alpha, \veck_\ee^\beta} =
    \sum_{\veck'\vecq'}v_{q'}
    \phi^*
    c_{-\veck' - \vecq'}^{\alpha \dagger}d_{\veck_\xx + \veck'}^\dagger
    c_{\veck_\ee + \vecq'}^{\beta \dagger}\ket{0}.
\end{equation}

From here by direct computation we find the general matrix elements of the
electron-exciton elastic scattering process to be
\begin{multline}
    \mel{(\veck_\xx + \vecq)^\theta, (\veck_\ee - \vecq)^\omega}{V_{\ee\hh}}{\veck_\xx^\alpha, \veck_\ee^\beta} = \\
    - \sum_{\veck'' \veck' \vecq'}\phi_1 \phi_2^* v_{q'}
    (\delta_{\veck'', \veck'}^{\theta \alpha}
    \delta_{-\vecq, \vecq'}^{\omega \beta}
     - \delta_{-\veck'', \veck_\ee +\vecq'}^{\theta \beta}
     \delta_{\veck', \veck_\ee - \vecq}^{\omega \alpha}),
\end{multline}
where $\phi_1\phi_2^* \equiv \phi_{\alpha_\xx \veck_\xx + \alpha_\xx \vecq + \veck''}
\phi_{\alpha_\xx \veck_\xx + \veck'}^*.$ Explicitly, this is
\begin{multline}
    \mel{(\veck_\xx + \vecq)^\theta, (\veck_\ee - \vecq)^\omega}{V_{\ee\hh}}{\veck_x^\alpha, \veck_e^\beta} = \\
    v_q \delta_{\theta \alpha}
    \delta_{\omega \beta}
    \sum_{\veck'} \phi_{\alpha_\xx \veck_\xx + \alpha_\xx \vecq 
    + \veck'}\phi_{\alpha_\xx \veck_\xx + \veck'} \\
    - \phi_{\alpha_\xx \veck_\xx - \veck_\ee + \vecq} \delta_{\theta \beta}
    \delta_{\omega \alpha}
    \sum_{\veck'} \phi_{\alpha_\xx \veck_\xx + \alpha_\xx \vecq - \veck_\ee
    - \veck'} v_{k'},
\end{multline}
which can be separated into direct (corresponding to $v_q$) and
exchange ($v_{k'}$) contributions. It is also observed that for practical
computation $\phi = \phi^*$ and thus the complex conjugation is dropped.
The electron-electron term is computed
\begin{multline}
    \mel{(\veck_\xx + \vecq)^\theta, (\veck_\ee - \vecq)^\omega}{V_{\ee\ee}}{\veck_\xx^\alpha, \veck_\ee^\beta }= \\
    \sum_{\veck'' \veck' \vecq'}\phi_1 \phi_2^* v_{q'}
    (\delta_{\veck'', \veck' + \vecq'}^{\theta \alpha}
    \delta_{\vecq, -\vecq'}^{\omega \beta}\\
     - \delta_{-\veck'', \veck_\ee +\vecq'}^{\theta \beta}
     \delta_{-\veck' - \vecq', \veck_\ee - \vecq}^{\omega \alpha})
\end{multline}
and simplified in a similar fashion,
\begin{multline}
    \mel{(\veck_\xx + \vecq)^\theta, (\veck_\ee - \vecq)^\omega}{V_{\ee\ee}}{\veck_\xx^\alpha, \veck_\ee^\beta} = \\
    v_q \delta_{\theta \alpha}
    \delta_{\omega \beta}
    \sum_{\veck'} \phi_{\alpha_\xx \veck_\xx - \beta_\xx \vecq 
    + \veck'}\phi_{\alpha_\xx \veck_\xx + \veck'} \\
    -  \delta_{\theta \beta}
    \delta_{\omega \alpha}
    \sum_{\veck'} \phi_{\alpha_\xx \veck_\xx + \alpha_\xx \vecq - \veck_\ee - \veck'}
    \phi_{\alpha_\xx\veck_\xx + \vecq - \veck_\ee - \veck'}  v_{k'}.
\end{multline}
Combining terms into direct and exchange contributions, we have
\begin{multline} \label{V direct  1s (appendix)}
    V^\mathrm{D}(\vecq, \veck_\ee, \veck_\xx) = v_q\delta_{\theta \alpha}
    \delta_{\omega \beta} \sum_{\veck'} 
    \phi_{\alpha_\xx \veck_\xx + \veck'} \\
     \times [\phi_{\alpha_\xx \veck_\xx - \beta_\xx \vecq + \veck'}
    - \phi_{\alpha_\xx \veck_\xx + \alpha_\xx \vecq + \veck'}],
\end{multline}
where the Kronecker delta functions ensure the proper spins are paired, and
\begin{multline} \label{V exchange  1s (appendix)}
    V^{\mathrm{XC}}(\vecq, \veck_\ee, \veck_\xx) = -\delta_{\theta \beta}
    \delta_{\omega \alpha} \sum_{\veck'} v_{k'} 
    \phi_{\alpha_x \vecq -\Delta \veck_\xx + \veck'}  \\
    \times [
    \phi_{\vecq - \Delta \veck_\xx + \veck'} - \phi_{\vecq - \Delta \veck_\xx}],
\end{multline}
where $\Delta \veck_\xx \equiv \veck_\ee - \alpha_\xx \veck_\xx.$

\subsection{Spin contributions}
Both the $V_{\ee\ee}$ and $V_{\ee\hh}$ terms can
be split into clear direct and exchange contributions such that in the
individual electron spin basis, 
$$\mel{\theta \omega}{V}{\alpha \beta} = 
\delta_{\theta \alpha}\delta_{\omega \beta} V^\mathrm{D} + \delta_{\theta \beta}
\delta_{\omega \alpha} V^\mathrm{XC}.$$

If the incident and exciton electrons are in a singlet configuration, we have
to consider all contributions from the singlet state 
$\ket{S} = (\ket{\uparrow \downarrow} - \ket{\downarrow\uparrow})/\sqrt{2},$
$$\mel{S}{V}{S} = \frac{1}{2} (\mel{\uparrow \downarrow}{V}{\uparrow\downarrow}
+ \text{cc.} -\mel{\uparrow \downarrow}{V}{\downarrow\uparrow} -
\text{cc.}),$$
which in the specified basis is
$$V_S \equiv \mel{S}{V}{S} = V^\mathrm{D} - V^\mathrm{XC}.$$
By inspection, any of the triplet configurations are
$$V_T \equiv \mel{T}{V}{T} = V^\mathrm{D} + V^\mathrm{XC}.$$
In the case of the exciton case, the singlet and triplet contributions are essentially
identical, since the exchange contribution dominates, meaning 
$\abs{V_S}^2 \approx \abs{V_T}^2;$ for the trion, we do not consider triplet states.

\subsection{1s$\rightarrow$1s scattering}
With the assumption that the exciton wave
function $\phi$ is in the parameterized ground state (1s) given by
(\ref{exciton wave function (k space)}), the direct interaction has an
analytic form. Noting that
$$ \sum_{\veck} \rightarrow \frac{A}{(2\pi)^2} \int \dd^2k, \quad 
\veck \in \mathbb{R}^2,$$
and that the convolution
\begin{equation} \label{convolution}
\int \dd^2k'g(\lambda k')g(\lambda'|\vecq \pm \veck'|) =
\frac{2\pi}{(\lambda + \lambda')^2}g\left(\frac{\lambda\lambda'q}{\lambda + \lambda'} \right),
\end{equation}
the direct terms simplify to (dropping the spin Kronecker deltas)
\begin{equation}
    V^\mathrm{D}_\mathrm{1s}(q)=
    \frac{2\pi e^2}{Aq \varepsilon(q)} \left[g(\lambda \beta_\xx q/2) - g(\lambda \alpha_\xx q/2) \right].
\end{equation}
The exchange terms do not simplify and must be evaluated numerically,
\begin{multline}
    V^\mathrm{XC}_{1s}(\vecq, \veck_\ee, \veck_\xx) = -\frac{4 e^2 \lambda^2}{A} \int \frac{\dd^2k'}{k'\varepsilon(k')} \\
    \times g(\lambda|\alpha_\xx \vecq - \Delta \veck_\xx +
    \veck'|) \\
    \times \left[g(\lambda|\veck_1 + \vecq - \Delta \veck_\xx|)
    - g(\lambda|\vecq - \Delta \veck_\xx|) \right].
\end{multline}
These results have been previously derived for scattering in finite quantum
wells\cite{Cohen2003} and match the results above in the 2D analytic limit. Here,
$\lambda = \lambda_0 = 10.3$~\AA~is the exciton effective Bohr radius, and $\alpha_\xx = m_\ee/M_\xx$ is
the mass ratio of the exciton, $\beta_\xx = 1 - \alpha_\xx,$ $m_\ee = 0.49 m_0$ and $M_\xx = m_\ee + m_\hh,$ where $m_\hh = 0.61 m_0.$

\subsection{2s$\rightarrow$2s scattering} \label{2s apdx}
To compute the excited state (2s) exciton elastic scattering line width,
we parameterize a radial 2s hydrogen wave function
\begin{equation}
    \phi^\mathrm{2s}(r, a, b) \propto \left( 2 - \frac{r}{b}\right)e^{-r/2a}
\end{equation}
in terms of an effective Bohr radius $a$ and a secondary parameter $b$ chosen to ensure
orthogonality to the 1s state. An initial fit to the first excited state
exact-diagonalization result of the Wannier exciton in a 2D Keldysh potential yielded
length scales $a = 7.79$~\AA~and $b=5.33$~\AA, the latter of which was modified to
$b=6.20$~\AA~to ensure orthogonality. Fourier transforming to momentum-space yields
\begin{equation}\label{2s exciton wave function}
    \phi_k^\mathrm{2s} = N_\mathrm{2s} \left[ \frac{16\pi a^2}{(1 + 4a^2k^2)^{3/2}}
    - \frac{2\pi \left( \frac{1}{2a^2} - k^2 \right)}{b \left(\frac{1}{4a^2} + k^2 \right)^{5/2}} \right]
\end{equation}
with normalization
$$ N_\mathrm{2s} = \sqrt{\frac{b^2}{4\pi a^2A(3a^2-4ab+2b^2)}}.$$

Matrix elements are computed by making the substitution $\phi \rightarrow \phi^\mathrm{2s}$ in
(\ref{V direct  1s (appendix)}) and (\ref{V exchange  1s (appendix)}) numerically performing the
2D integrals.

\section{Trion-electron elastic scattering} \label{apdx trion}
The details of the $T + e^- \rightarrow T + e^-$ scattering process are
significantly more involved than the exciton case. The introduction of an extra electron
manifests as another pair of creation and annihilation operators in the matrix element
evaluation and adds many more terms. While the calculation is longer, it is
no more conceptually difficult. In this appendix, we present the detailed
derivation of the matrix elements for a two dimensional system, which coincide with the results
for the 3D quantum well in the $L\rightarrow 0$ limit.\cite{Cohen2003}

The total elastic scattering matrix element $\mathcal{V}(\vecq, \veck_\ee, \veck_\tri)$
is calculated by first computing the action of $V | \veck_\tri, \veck_\ee \rangle.$ This
not only simplifies the number of operator contractions, it also allows for removal of
self-interaction terms (those characterized by internal interactions between electrons and
holes within the trion), as they do not contribute to the scattering matrix elements (similar to
that of the exciton scattering case). To begin, we first evaluate the general contraction,
which is used during the evaluation of (\ref{trion mel}),
\begin{equation}
    c_1c_2^\dagger c_3^\dagger c_4^\dagger\ket{0} 
    = [\delta_{12} c_3^\dagger c_4^\dagger -\delta_{13} c_2^\dagger c_4^\dagger + \delta_{14} c_2^\dagger c_3^\dagger]\ket{0},
\end{equation}
where in (\ref{Veh on trion derivation 1}),
$1 \equiv (\veck_1', z_1', s'),$
$2 \equiv (-\veck_1, z_1, s_1),$
$3 \equiv (-\veck_2, z_2, s_2),$ and
$4 \equiv (\veck_\ee, z_\ee, s_\ee),$
as this will be useful in computing both $V_{\ee\hh}|\veck_\tri, \veck_\ee \rangle$ and 
$V_{\ee\ee}|\veck_\tri, \veck_\ee \rangle.$ In following calculations, hole operators will
be ignored, as they do not contribute additional constraints or prefactors to the line width
calculations. Moreover, $V_{\ee\hh}|\veck_\tri, \veck_\ee \rangle$ evaluates to
\begin{widetext}
    \begin{equation} \label{Veh on trion derivation 1}
        \begin{split}
    V_{\ee\hh}\ket{\veck_\tri, \veck_\ee} &= -\sum_{\substack{\veck_1, \veck_2, \veck_1',  \vecq' \\ s_1, s_2, s_\ee, s'}} v_{q'} \xi_S^*(s_1, s_2)\Phi_{\alpha_\tri\veck_\tri + \veck_1,\alpha_t\veck_\tri + \veck_2}^* \psi_{\veck_\ee}^*
    c_{\veck_1'+\vecq'}^{s'\dagger}c_{\veck_1'}^{s'}c_{-\veck_1}^{s_1\dagger}
    c_{-\veck_2}^{s_2\dagger}c_{\veck_e}^{s_e\dagger}\ket{0}\\
    &= -\sum_{\substack{\veck_1, \veck_2, \vecq' \\ s_1, s_2, s_e}}
    \xi_S^*(s_1, s_2)\Phi_{\alpha_\tri\veck_\tri + \veck_1,\alpha_\tri\veck_\tri + \veck_2}^*\psi_{\veck_\ee}^* v_{q'} 
    \left\{ c_{-\veck_1+\vecq'}^{s_1\dagger} c_{-\veck_2}^{s_2\dagger} c_{\veck_\ee}^{s_\ee\dagger}
    -c_{-\veck_2+\vecq'}^{s_2\dagger}c_{-\veck_1}^{s_1\dagger} 
    c_{\veck_\ee}^{s_\ee\dagger}
    +c_{\veck_e+\vecq'}^{s_\ee\dagger}c_{-\veck_1}^{s_1\dagger} c_{-\veck_2}^{s_2\dagger}\right\}\ket{0}.
        \end{split}
    \end{equation}
\end{widetext}
The first two terms correspond to self-interactions between the internal electrons and holes of the
trion. This is most easily seen by observing that after the action of $V_{\ee\hh}$ on the trion-free
electron state, the initial incident electron momentum $\veck_\ee$ remains unchanged in the final
creation operator. In the last term, however, we see that a momentum exchange of $\vecq'$ has taken
place.

The electron-electron terms corresponding to $V_{\ee\ee}\ket{\veck_\tri, \veck_\ee}$ are calculated
similarly. As in the electron-hole case, we begin by performing the
right-most contraction
\begin{equation} \label{contraction1}
\begin{split}
    c_1c_2c_3^\dagger c_4^\dagger c_5^\dagger \ket{0} &= [\delta_{15}\delta_{24}-\delta_{14}\delta_{25}]c_3^\dagger\ket{0}\\
    &+ [\delta_{13}\delta_{25}-\delta_{15}\delta_{23}]c_4^\dagger\ket{0}\\
    &+ [\delta_{14}\delta_{23}-\delta_{13}\delta_{24}]c_5^\dagger\ket{0},
\end{split}
\end{equation}
where in (\ref{contraction1}) and (\ref{Vee on trion derivation 1})
$1 \equiv(\veck_2', s_2'),$
$2 \equiv(\veck_1', s_1'),$
$3 \equiv(-\veck_1, s_1),$
$4 \equiv(-\veck_2, s_2),$
$5 \equiv (\veck_\ee, s_\ee).$
In similar fashion, $V_{\ee\ee}\ket{\veck_\tri, \veck_\ee}$ is thus found to be
\begin{widetext}
    \begin{equation}\label{Vee on trion derivation 1}
    \begin{split}
    V_{\ee\ee}|\veck_\tri, \veck_\ee\rangle &= 
    \frac{1}{2}\sum_{\substack{\veck_1, \veck_2, \veck_1', \veck_2', \vecq' \\
    s_1, s_2, s_\ee, s_1', s_2'}} 
    \xi_S^*(s_1, s_2)\Phi_{\alpha_\tri\veck_\tri + \veck_1,\alpha_\tri\veck_\tri + \veck_2}^* \psi_{\veck_\ee}^* v_{q'} c_{\veck_1'+\vecq'}^{s_1'\dagger}
    c_{\veck_2'-\vecq'}^{s_2'\dagger}c_{\veck_2'}^{s_2'}c_{\veck_1'}^{s_1'}
    c_{-\veck_1}^{s_1\dagger}c_{-\veck_2}^{s_2\dagger}
    c_{\veck_\ee}^{s_\ee\dagger}\ket{0} \\
    &=\frac{1}{2}\sum_{\substack{\veck_1, \veck_2, \vecq' \\
    s_1, s_2, s_\ee}} 
    \xi_S^*(s_1, s_2)\Phi_{\alpha_\tri\veck_\tri + \veck_1,\alpha_\tri\veck_\tri + \veck_2}^*\psi_{\veck_\ee}^* v_{q'} \big\{c_{-\veck_2+\vecq'}^{s_2\dagger}
    c_{\veck_\ee-\vecq'}^{s_\ee\dagger}c_{-\veck_1}^{s_1\dagger}
    - c_{\veck_\ee+\vecq'}^{s_\ee\dagger}
    c_{-\veck_2-\vecq'}^{s_2\dagger}c_{-\veck_1}^{s_1\dagger} \\
    &+ c_{\veck_\ee+\vecq'}^{s_\ee\dagger}
    c_{-\veck_1-\vecq'}^{s_1\dagger}c_{-\veck_2}^{s_2\dagger}
    - c_{-\veck_1+\vecq'}^{s_1\dagger}
    c_{\veck_\ee-\vecq'}^{s_\ee\dagger}c_{-\veck_2}^{s_2\dagger}
    +c_{-\veck_1+\vecq'}^{s_1\dagger}
    c_{-\veck_2-\vecq'}^{s_2\dagger}c_{\veck_\ee}^{s_\ee\dagger}
    -c_{-\veck_2+\vecq'}^{s_2\dagger}
    c_{-\veck_1-\vecq'}^{s_1\dagger}c_{\veck_\ee}^{s_\ee\dagger}\big\}\ket{0} \\
    &=\sum_{\substack{\veck_1, \veck_2, \vecq' \\
    s_1, s_2, s_e}}
    \xi_S^*(s_1, s_2)\Phi_{\alpha_\tri\veck_\tri + \veck_1,\alpha_\tri\veck_\tri + \veck_2}^* \psi_{\veck_\ee}^* v_{q'} \\
    &\times \big\{c_{-\veck_1}^{s_1\dagger}c_{-\veck_2-\vecq'}^{s_2\dagger}c_{\veck_\ee+\vecq'}^{s_\ee\dagger} + c_{-\veck_1-\vecq'}^{s_1\dagger}c_{-\veck_2}^{s_2\dagger}c_{\veck_\ee+\vecq'}^{s_\ee\dagger}
    +c_{-\veck_1-\vecq'}^{s_1\dagger}c_{-\veck_2+\vecq'}^{s_2\dagger}c_{\veck_\ee}^{s_\ee\dagger}\big\} \ket{0}
    \end{split}
    \end{equation}
\end{widetext}
In order to move from the second to the third equality in
(\ref{Vee on trion derivation 1}), we have made use of the substitutions $\vecq'
\rightarrow -\vecq'$ in the first, third and fifth terms. The last term
is a self-interaction exchange of momentum $\vecq'$ between the two electrons on the trion.

With the self-interaction terms removed, and ignoring hole operators, the
operation of $V = V_{\ee\hh} + V_{\ee\ee}$ acting on the trion-electron scattering
state is
\begin{multline}\label{V on trion-free electron}
    V\ket{\veck_\tri, \veck_\ee} = \sum_{\substack{\veck_1, \veck_2, \vecq' \\
    s_1, s_2, s_\ee}} \xi_S^*(s_1, s_2) 
    \Phi^*_{\alpha_\tri\veck_\tri + \veck_1, \alpha_\tri\veck_\tri + \veck_2}
    \psi_{\veck_\ee}^* v_{q'}\\ \times
    \big\{c_{-\veck_1-\vecq'}^{s_1\dagger}c_{-\veck_2}^{s_2\dagger}+
    c_{-\veck_1}^{s_1\dagger}c_{-\veck_2-\vecq'}^{s_2\dagger}
    -c_{-\veck_1}^{s_1\dagger}(z_1)c_{-\veck_2}^{s_2\dagger}\big\}c_{\veck_\ee +
    \vecq'}^{s_\ee\dagger}\ket{0}.
\end{multline}

The trion-electron elastic scattering matrix elements
are given by the action of $\bra{\veck_\tri + \vecq, \veck_\ee - \vecq}$ on (\ref{V on
trion-free electron}). Executing all possible integrals analytically produces a series
of terms which can be broken into a direct component and two
exchange components. We first adopt some notation:
$\widetilde \lambda = \lambda_1 \lambda_2/(\lambda_1 + \lambda_2),$
is a harmonic-mean-like term which arises during convolutions 
e.g. in (\ref{convolution}), and
$\Delta \veck_\tri = \veck_\ee - \alpha_\tri\veck_\tri,$
where $\veck_\tri$ is the initial trion momentum and $\alpha_\tri = m_\ee/M_\tri$ is the
ratio of the effective electron mass to that of the trion's $M_\tri = 2m_\ee +
m_\hh.$ Finally, the elastic scattering matrix elements,
$\mathcal{V}(\vecq, \veck_\ee, \veck_\tri),$
are given by the sum of (\ref{elastic direct terms}) and (\ref{elastic exchange
terms}). The direct terms are
\begin{equation} \label{elastic direct terms}
    \mathcal{V}^\text{D}(q) = \frac{2\pi e^2}{q\varepsilon(q) A(1+\kappa^2)}
\sum_{j=1}^5 g_j(q),
\end{equation}
where
\begin{equation} \label{elastic direct terms explicit}
\begin{aligned}
g_1(q) &= g(\lambda_1 \alpha_\tri q/2)g(\lambda_2\beta_\tri q/2), \\
g_2(q) &= g(\lambda_2 \alpha_\tri q/2)g(\lambda_1\beta_\tri q/2), \\
g_3(q) &= 2\kappa^2g(\widetilde \lambda \alpha_\tri q)
g(\widetilde \lambda \beta_\tri q), \\
g_4(q) &= - g(\lambda_1\alpha_\tri q/2)g(\lambda_2\alpha_\tri q/2), \\
g_5(q) &= -\kappa^2 g^2(\widetilde \lambda \alpha_\tri q),
\end{aligned}
\end{equation}
and the exchange terms are
\begin{multline} \label{elastic exchange terms}
    \mathcal{V}^\text{XC}(\vecq, \veck_\ee, \veck_\tri) = \frac{2 e^2}{A(1+\kappa^2)}
    \int \frac{\dd^2k'}{k'\varepsilon(k')}
     \\ \times \sum_{j=1}^6 
     \left[G_j(\vecq, \veck'; \lambda_1, \lambda_2)
    + G_j(\vecq, \veck'; \lambda_2, \lambda_1) \right],
\end{multline}
where
\begin{equation} \label{elastic exchange terms explicit}
\begin{aligned}
G_1(\vecq, \veck'; \lambda_1, \lambda_2)
&= \lambda_2^2g(\lambda_1\alpha_\tri q/2)g(\lambda_2|\Delta \veck_\tri - \vecq|) \\
&\times g(\lambda_2|\alpha_\tri \vecq - \Delta \veck_\tri + \veck'|),\\
G_2(\vecq, \veck'; \lambda_1, \lambda_2)
&= -\lambda_2^2 g(\lambda_2|\alpha_\tri \vecq - \Delta \veck_\tri + \veck'|) \\
&\times g(\lambda_1|\alpha_\tri \vecq - \veck'|/2)g(\lambda_2|\Delta \veck_\tri - \vecq|),  \\
G_3(\vecq, \veck'; \lambda_1, \lambda_2)
&= -\lambda_2^2 g(\lambda_1 \alpha_\tri q/2)g(\lambda_2|\vecq - \Delta \veck_\tri + \veck'|) \\
&\times g(\lambda_2|\alpha_\tri \vecq - \Delta \veck_\tri + \veck'|), \\
G_4(\vecq, \veck'; \lambda_1, \lambda_2)
&= \kappa \lambda_1 \lambda_2 g(\widetilde \lambda \alpha_\tri q)g(\lambda_1|\alpha_\tri \vecq - \Delta \veck_\tri + \veck'|) \\
&\times g(\lambda_2|\Delta \veck_\tri - \vecq|), \\
G_5(\vecq, \veck'; \lambda_1, \lambda_2)
&= - \kappa \lambda_1 \lambda_2 
g(\lambda_1|\alpha_\tri \vecq - \Delta \veck_\tri + \veck'|) \\
&\times g(\widetilde \lambda | \alpha_\tri \vecq - \veck'|)g(\lambda_2|\Delta \veck_\tri - \vecq|), \\
G_6(\vecq, \veck'; \lambda_1, \lambda_2)
&= - \kappa \lambda_1 \lambda_2 g(\widetilde \lambda \alpha_\tri q)
g(\lambda_2|\vecq - \Delta \veck_\tri + \veck'|)\\
&\times g(\lambda_1|\alpha_\tri \vecq - \Delta \veck_\tri + \veck'|).
\end{aligned}
\end{equation}

Applying $\bra{\veck_\tri + \vecq, \veck_\ee - \vecq}$
on (\ref{V on trion-free electron}) produces a series of integrals, many of
which may be evaluated analytically. In addition, the signs, and in some cases the
prefactor, of the various terms are determined by summing over the spin degrees of
freedom.

As in previous calculations, it is helpful to evaluate the contraction of
Fermionic operators
\begin{equation} \label{full contraction 6}
\begin{split}
\mel{0}{c_1 c_2 c_3 c_4^\dagger c_5^\dagger c_6^\dagger}{0}
&= \delta_{16}(\delta_{34}\delta_{25} - \delta_{24}\delta_{35}) \\
&+ \delta_{15}(\delta_{24}\delta_{36} - \delta_{34}\delta_{26}) \\
&+ \delta_{14}(\delta_{35}\delta_{26} - \delta_{25}\delta_{36}),
\end{split}
\end{equation}
where for the electron-hole interaction,
\begin{equation} \label{scattering matrix elements contraction labels}
\begin{aligned}
1 &\equiv (\veck_\ee-\vecq, s_\ee'), \\
2 &\equiv (-\veck_2', s_2'), \\
3 &\equiv (-\veck_1', s_1'), \\
4 &\equiv (-\veck_1, s_1), \\
5 &\equiv (-\veck_2, s_2), \\
6 &\equiv (\veck_\ee + \vecq', s_\ee).
\end{aligned}
\end{equation}
As an example, the terms including $\delta_{16}$ produce $g_4(q)$ and $g_5(q)$ in
(\ref{elastic direct terms explicit}), and the remainder of the exchange terms correspond to $G_1$ and
$G_4$ in (\ref{elastic exchange terms explicit}).

To evaluate the first (second) electron-electron interactions, we replace
$\veck_1 \rightarrow \veck_1 + \vecq'$ ($\veck_2 \rightarrow \veck_2 + \vecq'$) in 
(\ref{scattering matrix elements contraction labels}). Each of the six terms
generated by the contraction in (\ref{full contraction 6}) is evaluated
individually for the hole and two electrons, generating 18 total terms and 
producing (\ref{elastic direct terms explicit}) and (\ref{elastic exchange
terms explicit}). Note that $g_3(q)$ in (\ref{elastic direct terms explicit})
accounts for two identical electron-electron interaction terms.

Instead of presenting a derivation of all 18 terms, we present a detailed derivation of one of them. The others follow similarly. Consider the electron-hole term corresponding to $\delta_{24}\delta_{15}\delta_{36},$
\begin{multline}
-\frac{1}{2} \sum_{\substack{\veck_1, \veck_2, \veck_1', \veck_2', \vecq' \\
s_1, s_2, s_e, s_1', s_2', s_\ee'}}v_{q'} \xi_S^*(s_1, s_2) \xi_S(s_1', s_2')\psi_{\veck_\ee}^*\psi_{\veck_\ee-\vecq} \\
\times \Phi^*_{\alpha_\tri\veck_\tri + \veck_1, \alpha_\tri\veck_\tri + \veck_2}
\Phi_{\alpha_\tri(\veck_\tri+\vecq) + \veck_1', \alpha_\tri(\veck_\tri+\vecq) + \veck_2'}\\
\times \delta_{\veck_2',\veck_1}\delta_{s_2', s_1}
\delta_{\veck_\ee - \vecq, -\veck_2}\delta_{s_\ee', s_2}
\delta_{-\veck_1', \veck_\ee + \vecq'}\delta_{s_1', s_\ee}.
\end{multline}
Note that the factor of $1/2$ is due to an average
over the initial free-electron spin states. We may sum  over the
dummy variables $s_1', s_2', s_\ee', \veck_2', \veck_2$ and $\vecq'.$ This yields
\begin{multline} \label{Vprime 2}
-\frac{1}{2}\sum_{\substack{\veck_1, \veck_1'\\
s_1, s_2, s_\ee}} v_{\veck_\ee + \veck_1'} \xi_S^*(s_1, s_2)
\xi_S(s_e, s_1)\psi_{\veck_\ee}^*\psi_{\veck_\ee-\vecq} \\
\times \Phi^*_{\alpha_\tri\veck_\tri + \veck_1, \alpha_\tri\veck_\tri + \vecq - \veck_\ee}
\Phi_{\alpha_\tri(\veck_\tri+\vecq) + \veck_1', \alpha_\tri(\veck_\tri+\vecq) + \veck_1}.
\end{multline}
The spin factors are evaluated first:
\begin{equation}
    \sum_{s_1, s_2, s_e} \xi_S^*(s_1, s_2) \xi_S(s_e, s_1) = 
    \begin{cases}
    -1 & \text{if } s_2 = s_e \\
    0 & \text{if } s_2 \neq s_e.
    \end{cases}
\end{equation}
After sorting each term in (\ref{Vprime 2} by integration variable and 
making the substitutions $\veck_1' \rightarrow \veck_1' - \veck_\ee$
and $\veck_1 \rightarrow \veck_1 - \alpha_\tri \veck_\tri,$
one integral may be evaluated analytically using the convolution in (\ref{convolution}), yielding
$G_4(\vecq, \veck'; \lambda_1, \lambda_2)$ after the
substitution $\Delta \veck_\tri = \veck_\ee - \alpha_\tri \veck_\tri$ is made.

\section{Computational Details \label{apdx: computational}}

All integrations were performed using the Cubature adaptive
integration package.\cite{hcube} Integrals over $(0, \infty)$
were mapped to the finite range $(0, 1)$ and performed using adaptive integration.
Additionally, in order to avoid exhausting available
memory, integrals were nested in the following way: First, scattering
matrix elements were computed on the fly and converged to some
relative error tolerance $\epsilon.$ This results in a computation of
a two-dimensional integral for the exchange terms. Once the matrix
element $V$ is computed, the Golden Rule integration which contains
$|V|^2$, along with
the integration over all final states, is performed (this is a 
three-dimensional integral), and converged to some error tolerance
$c\epsilon,$ where $c$ is typically on the order of $100-1000.$
Finally, to ensure convergence, the entire computation is converged
with respect to the decreasing of $\epsilon.$


%

\end{document}